\documentstyle[prl,multicol,aps,epsfig]{revtex}
\renewcommand{\widetext}
{\end{multicols}\global\columnwidth42.5pc}
\begin{document}
\newcommand{\be}{\begin{equation}}
\newcommand{\ee}{\end{equation}}
\newcommand{\bea}{\begin{eqnarray}}
\newcommand{\eea}{\end{eqnarray}}
\newcommand{\br}{{\bf r}}
\newcommand{\bk}{{\bf k}}
\newcommand{\bq}{{\bf q}}
\newcommand{\bn}{{\bf n}}
\newcommand{\bp}{{\bf p}}
\newcommand{\bE}{{\bf E}}
\newcommand{\ve}{\varepsilon}

\draft
\title{Theory of the oscillatory photoconductivity of a 2D electron gas}
\author{I.A.~Dmitriev$^{1,*}$, M.G.~Vavilov$^{2}$, I.L.~Aleiner$^{3}$,
A.D.~Mirlin$^{1,4,\dagger}$, and D.G.~Polyakov$^{1,*}$}
\address{$^1$Institut f\"ur Nanotechnologie, Forschungszentrum
Karlsruhe, 76021 Karlsruhe, Germany}
\address{$^2$ Department of Physics,
Massachusetts Institute of Technology,
Cambridge, MA 02139, USA}
\address{$^3$Physics Department, Columbia University, New York, NY
10027, USA}
\address{$^4$Institut f\"ur Theorie der Kondensierten Materie,
Universit\"at Karlsruhe, 76128 Karlsruhe, Germany}
\maketitle
\begin{abstract}
We develop a theory of magnetooscillations in the photoconductivity
of a two-dimensional electron gas observed in recent experiments.
The effect is governed by a
change of the electron distribution function induced by the
microwave radiation. We analyze a nonlinearity with respect to both
the {\it dc} field and the microwave power, as well as the temperature
dependence determined by the inelastic relaxation rate.
\end{abstract}
\begin{multicols}{2}
\narrowtext

Recent experiments have discovered \cite{zudov01} that the resistivity
of a high-mobility two-dimensional electron gas (2DEG) in GaAs/AlGaAs
heterostructures subjected to microwave radiation of frequency
$\omega$ exhibits magnetooscillations governed by the ratio
$\omega/\omega_c$, where $\omega_c$ is the cyclotron frequency.
Subsequent work \cite{mani02,zudov03,yang03,dorozhkin03,willett03} has shown
that for samples with a very high mobility and for 
high radiation power the minima of the oscillations
evolve into zero-resistance states (ZRS).

These spectacular observations
have attracted much theoretical interest.
As was shown in Ref.~\cite{andreev03},
the ZRS can be understood as a direct consequence of the
oscillatory photoconductivity (OPC), provided that
the latter may become negative.
A negative value of the OPC 
signifies an instability leading to the formation of
spontaneous-current domains showing zero value of the
observable resistance. Therefore,
the identification of the microscopic mechanism of the OPC appears
to be the key question in the interpretation of
the data \cite{zudov01,mani02,zudov03,yang03,dorozhkin03,willett03}.

A mechanism of the OPC proposed in Ref.~\cite{durst03}
is based on the effect of microwave radiation on
electron scattering by impurities in a strong magnetic field (see
also Ref.~\cite{ryzhii} for an earlier theory
and Ref.~\cite{vavilov03} for a systematic theory).
An alternative mechanism of the OPC was
recently proposed in Ref.~\cite{dmitriev03}. In contrast to
Refs.~\cite {durst03,ryzhii,vavilov03}, this mechanism is governed by
a radiation-induced change of the electron distribution function. Because
of the oscillations of the density of states (DOS), $\nu(\ve)$,
related to the Landau quantization, the correction to the distribution
function acquires an oscillatory structure as well. This
generates a contribution to the {\it dc} conductivity which oscillates
with varying $\omega/\omega_c$. A distinctive feature of the
contribution of Ref.~\cite{dmitriev03} is that it is
proportional to the inelastic relaxation time $\tau_{\rm in}$.  A
comparison of the results of Refs.~\cite{vavilov03} and \cite{dmitriev03}
shows that the latter contribution dominates if $\tau_{\rm in}\gg
\tau_{q}$ (where $\tau_q$ is the quantum, or single-particle,
relaxation time due to impurity scattering), which is the case for the
experimentally relevant temperatures.

The consideration of Ref.~\cite{dmitriev03} is restricted to the
regime which is linear in both the {\it ac} power and the {\it dc}
electric field. The purpose of this paper is to develop a complete
theory of the OPC governed by this
mechanism, including nonlinear effects. We will demonstrate that the
conductivity at a minimum becomes negative for a large
microwave power and that a positive sign is restored for a strong {\it
dc} bias, as it was assumed in Ref.~\cite{andreev03}.

We consider a 2DEG (mass $m$, density $n_e$, Fermi
velocity $v_F$) subjected to a transverse
magnetic field $B=(mc/e)\,\omega_c$. We assume that the field is
classically strong, $\omega_c\tau_{\rm tr}\gg 1$, where $\tau_{\rm
tr}$ is the transport relaxation time at $B=0$. The
photoconductivity $\sigma_{\rm ph}$ determines the longitudinal
current flowing in response to a {\it dc} electric field ${\cal
E}_{\rm dc}$, $\vec{j}\cdot\vec{\cal E}_{\rm dc}=\sigma_{\rm
ph}{\cal E}_{\rm dc}^2$, in the presence of a microwave electric field
${\bf{\cal E}}_\omega\cos\omega t$. The more
frequently   measured
\cite{zudov01,mani02,zudov03,dorozhkin03,willett03}
longitudinal resistivity,
$\rho_{\rm ph}$, is  given by
$\rho_{\rm ph}\simeq \rho_{xy}^2\sigma_{\rm ph}$, where
$\rho_{xy}\simeq eB/n_ec$ is the Hall resistivity, affected only
weakly by the radiation.

We start with the formula for the {\it dc}  conductivity:
\be
\label{photo}
\sigma_{\rm ph}=2 \int d\ve \,
\sigma_{\rm dc}(\ve)\, \left[-\partial_\ve
f(\ve)\right],
\ee
where $f(\ve)$ is the electron distribution function, and
$\sigma_{\rm dc}(\ve)$
determines the
contribution of electrons with energy $\ve$ to the dissipative
transport. In the leading approximation
\cite{vavilov03,dmitriev03},
$\sigma_{\rm dc}(\ve)=\sigma^{\rm
D}_{\rm dc}\,\tilde{\nu}^2(\ve)$, where
$\tilde{\nu}(\ve)=\nu(\ve)/\nu_0$ is the dimensionless DOS,
$\nu_0=m/2\pi$ is the DOS per spin at zero $B$
(we use $\hbar=1$), and
$\sigma^{\rm D}_{\rm dc}=e^2\nu_0 v_{\rm F}^2/2\omega_c^2\tau_{\rm tr}$
is the {\it dc} Drude conductivity per spin.
All interesting effects are due to a non-trivial energy
dependence of the non-equlibirum distribution function $f(\ve)$.
The latter is found as a solution of the stationary kinetic equation
\bea
\label{kineq}
\nonumber
&&{\cal E}^2_\omega\,\frac{\sigma^{\rm D}(\omega)}{2\omega^2\nu_0}
\sum\limits_{\pm}\tilde{\nu}(\ve\pm\omega)\,[\,f(\ve\pm\omega)-f(\ve)\,]\\
&&+\,\,{\cal E}^2_{\rm dc}\,\frac{\sigma^{\rm D}_{\rm
dc}}{\nu_0\tilde{\nu}(\ve)} \,\frac{\partial}{\partial\ve}
\left[\,\tilde{\nu}^2(\ve)\frac{\partial}{\partial\ve}f(\ve)\,\right]
={f(\ve)-f_T(\ve)\over\tau_{\rm in}},
\eea
where the {\it ac} Drude conductivity per spin is given by
(we assume $|\omega\pm\omega_c| {\tau_{\rm tr}}\gg 1$)
\be
\label{Drude}
\sigma^{\rm D}(\omega)=\sum_{\pm}\,\frac{e^2\nu_0 v_{\rm F}^2}
{4 \tau_{\rm tr}(\omega\pm\omega_c)^2}.
\ee
On the right-hand side of Eq.~(\ref{kineq}),
inelastic processes are included in the relaxation
time approximation (more detailed discussion of the relaxation
time $\tau_{\rm in}$ is relegated to the end of the paper),
and $f_T(\ve)$ is the Fermi distribution.
The left-hand side  is due to the electron collisions with
impurities in the presence of the external electric fields.
The first term describes the absorption and emission of
microwave quanta; the rate of these transitions was calculated in
Ref.~\cite{dmitriev03}. This term can be also extracted
from the kinetic equation of Ref.~\cite{vavilov03}.
The second term describes the effect of the
{\it dc} field and can be obtained from the first one by taking the
limit $\omega\to 0$.


Equation (\ref{kineq}) suggests convenient
dimensionless units for the
strength of the {\it ac} and {\it dc} fields:
\begin{mathletters}
\label{unitsPQ}
\bea
&&{\cal P}_\omega
=\frac{\tau_{\rm in}}{\tau_{\rm tr}}
\left(\frac{e {\cal E}_\omega v_F}{\omega}\right)^2
\frac{\omega_c^2+\omega^2}{(\omega^2-\omega_c^2)^2},
\label{units1}\\
&&{\cal Q}_{\rm dc}=\frac{2\,\tau_{\rm in}}{\tau_{\rm tr}}
\left(\frac{e {\cal E}_{\rm dc} v_F}{\omega_c}\right)^2
\left(\frac{\pi}{\omega_c}\right)^2.\label{units}
\eea
Note that ${\cal P}_\omega$ and ${\cal Q}_{\rm dc}$ are proportional
to $\tau_{\rm in}$ and are infinite in the absence of
inelastic relaxation processes.
\end{mathletters}

We consider first the case of overlapping Landau levels (LLs), with
the DOS given by $\tilde{\nu}=1- 2\delta\cos
\frac{2\pi\ve}{\omega_c}$,
where
$\delta=\exp(-\pi/\omega_c\tau_{\rm q})\ll 1$. Here $\tau_{\rm
  q}$ is the zero-$B$ single-particle relaxation time, which is much
shorter  than the transport time in high-mobility structures,
$\tau_{\rm q}\ll \tau_{\rm tr}$ (because of the smooth character of a
random potential of remote donors).  The existence of a small
parameter $\delta$  simplifies solution of the kinetic
equation (\ref{kineq}). To first order in $\delta$, we look for
a solution in the form
\be
\label{fe-expansion}
f=f_0+f_{\rm osc}+O(\delta^2), \ \ \ \ 
f_{\rm osc}\equiv \delta\,
{\rm Re}\left[f_1(\ve)\,e^{i\frac{2\pi\ve}{\omega_c}}\right].
\ee
We assume that the electric fields are not too strong [\,${\cal
P}_\omega(\omega/T)^2\ll 1$ and ${\cal Q}_{\rm dc}(\omega_c/T)^2\ll
1$\,], so that the smooth part $f_0(\ve)$ is close to the Fermi
distribution $f_T(\ve)$ at a bath temperature $T \gg \omega_c$;
otherwise, the temperature of the electron gas is further increased
due to heating. Smooth functions $f_{0,1}(\ve)$ change on a scale of
the order of temperature. We obtain
\be
\label{distr}
f_{\rm osc}(\ve) = \delta\,\frac{\omega_c}{2\pi}\:
\frac{\partial f_T}{\partial \ve}\:
\sin \frac{2\pi\ve}{\omega_c}\:
\frac{{\cal P}_\omega\frac{2\pi \omega}{\omega_c}
\sin\frac{2\pi\omega}{\omega_c}
 +4{\cal Q}_{\rm dc}} {1+{\cal P}_\omega
 \sin^2\frac{\pi\omega}{\omega_c}
+{\cal Q}_{\rm dc}}
\ee
and substitute Eq.~(\ref{distr})
into Eq.~(\ref{photo}).
Performing the energy integration in Eq.~(\ref{photo}),
we assume (in conformity with the experiment) that
$T$ is much larger than the Dingle temperature,
$T\gg 1/2\pi\tau_{q}$.
The terms of order $\delta$ in Eq.~(\ref{photo}) are
exponentially suppressed
$\delta \int
d\ve\, \partial_\ve f_T \cos \frac{2\pi\ve}{\omega_c}
\propto \delta\exp (-2\pi^2 T/\omega_c) \ll \delta^2$  and
can be
neglected. The leading $\omega$ dependent contribution to $\sigma_{\rm ph}$
 comes from the $\delta^2$ term generated
by the product of $\partial_\ve f_{\rm osc}(\ve) \propto
\delta\cos \frac{2\pi\ve}{\omega_c}$
and the oscillatory part $-2\delta \cos\frac{2\pi\ve}{\omega_c}$ of
$\tilde{\nu}(\ve)$. This term does survive the energy averaging, $-\int
d\ve\, \partial_\ve f_T \cos^2\frac{2\pi\ve}{\omega_c} \simeq 1/2$. We
thus find
\be
\label{result}
\frac{\sigma_{\rm ph}\ }{\sigma^{\rm D}_{\rm dc}}=1+2\delta^2\left[\,1
-
\frac{{\cal P}_\omega\frac{2\pi \omega}{\omega_c}
\sin\frac{2\pi\omega}{\omega_c}
 +4{\cal Q}_{\rm dc}} {1+{\cal P}_\omega
 \sin^2\frac{\pi\omega}{\omega_c}
+{\cal
Q}_{\rm dc}}
\right].
\ee

Equation (\ref{result}) is our central result. It
describes the photoconductivity in the regime of overlapping LLs,
including all non-linear (in ${\cal E}_\omega$ and ${\cal E}_{\rm
dc}$) effects.  Let us analyze it in more detail. In the
linear-response regime (${\cal E}_{\rm dc}\to 0$) and for a not too
strong microwave field, Eq.~(\ref{result}) yields a correction to the
dark {\it dc} conductivity $\sigma_{\rm dc}=
\sigma^{\rm D}_{\rm dc}(1+2\delta^2)$ which is linear in the microwave power:
\be
\label{linear}
{\sigma_{\rm ph}-\sigma_{\rm dc}\over\sigma_{\rm dc}} = - 4 \delta^2
{\cal P}_\omega \,{\pi\omega\over\omega_c} \,\sin
{2\pi\omega\over\omega_c},
\ee
in agreement with Ref.~\cite{dmitriev03}.
It is enlightening to compare Eq.~(\ref{linear}) with the
contribution of the effect of the {\it ac} field on the impurity scattering
\cite{durst03,ryzhii,vavilov03}. The
analytic result, Eq.~(6.11) of Ref.~\cite{vavilov03}, in the notation
of Eq.~(\ref{unitsPQ}) is
\[
{\sigma_{\rm ph}^{\cite{vavilov03}}-
\sigma_{\rm dc}\over\sigma_{\rm dc}} =
-  12 \frac{\tau_{{\rm q}} }{\tau_{\rm in}}
\delta^2 {\cal P}_\omega
\left({\pi\omega\over\omega_c} \,\sin
{2\pi\omega\over\omega_c} + \sin^2\frac{\pi\omega}{\omega_c}
\right).
\]
This result has a similar frequency dependence as
Eq.~(\ref{linear}); however, its amplitude is
much smaller at $\tau_{\rm in} \gg \tau_{{\rm q}}$, i.e., the  mechanism
of Refs.~\cite{durst03,ryzhii,vavilov03} appears to be irrelevant.
Physically, the effect of the {\it ac} field on the
distribution function is dominant
because it is accumulated during a diffusive process of duration
$\tau_{\rm in}$, whereas Refs.~\cite{durst03,ryzhii,vavilov03} consider
only one scattering event.

 \begin{figure}[ht]
 \narrowtext
 \centerline{ {\epsfxsize=7cm{\epsfbox{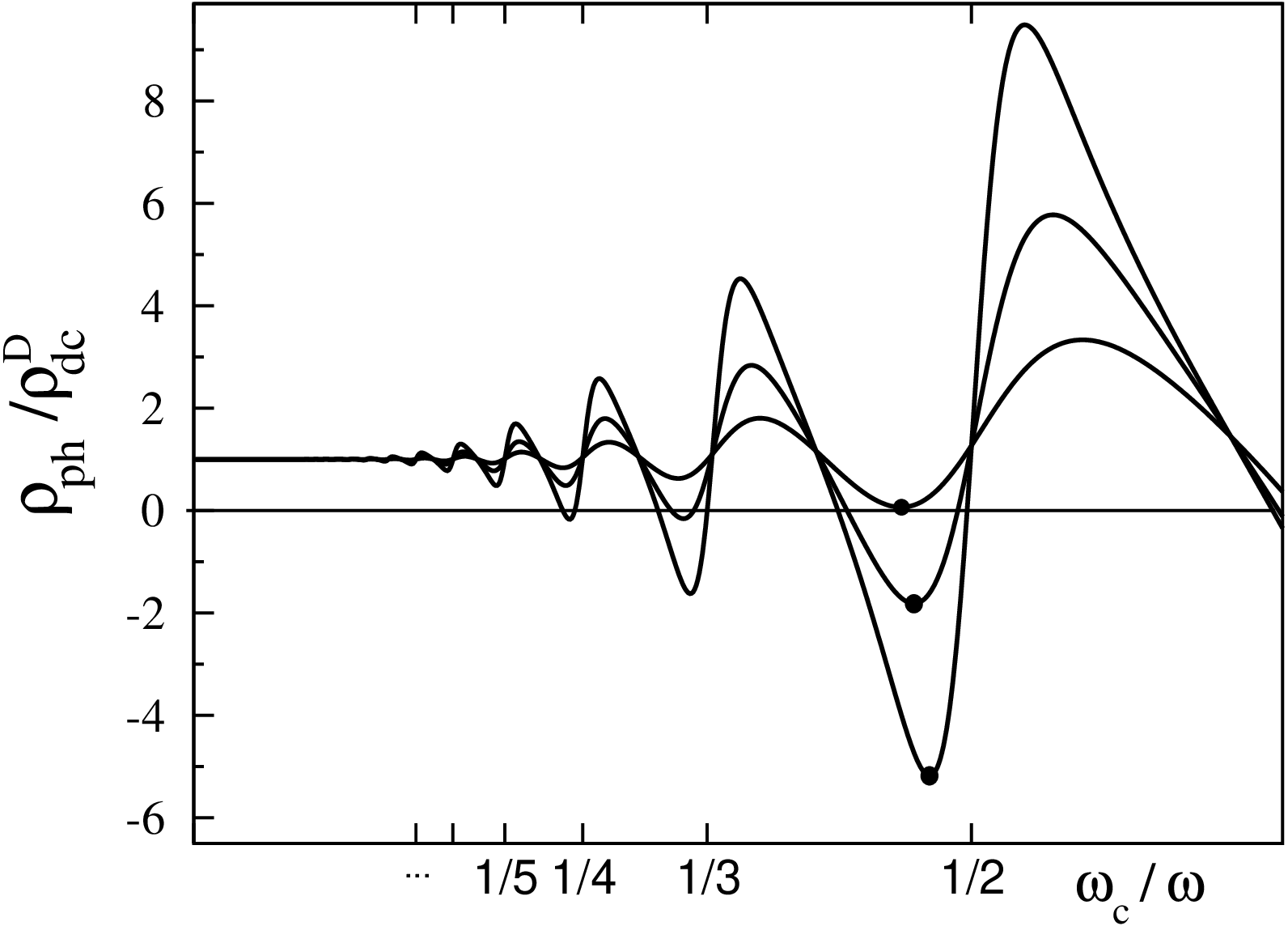}} }}
 \vspace{3mm}
 \caption{Photoresistivity (normalized to the dark Drude value) for
overlapping Landau levels vs $\omega_c/\omega$ at fixed
$\omega\tau_{\rm q}=2\pi$. The curves correspond to
different levels of microwave power ${\cal P}_\omega^{(0)}
=\{0.24,\,0.8,\,2.4\}$. Nonlinear $I-V$ characteristics at the marked
minima are shown in Fig.~\ref{fig2}.}
 \label{fig1}
 \end{figure}

With increasing microwave
power, the photoconductivity saturates at the value
\be
\label{saturation}
{\sigma_{\rm ph}\over\sigma_{\rm dc}} = 1- 8 \delta^2
\,{\pi\omega\over\omega_c} \,\cot {\pi\omega\over\omega_c}~, \quad
{\cal P}_\omega\sin^2 {\pi\omega\over\omega_c}\gg 1.
\ee
Note that although the correction is proportional to $\delta^2\ll 1$,
the factor $8\pi(\omega/\omega_c)\cot(\pi\omega/\omega_c)$ is large in
the vicinity of the cyclotron resonance harmonics $\omega=k\omega_c$
($k=1,2,\ldots$), and allows the photo-induced correction to exceed in
magnitude the dark conductivity $\sigma_{\rm dc}$.
In particular, $\sigma_{\rm ph}$ around minima becomes negative at
${\cal P}_\omega >{\cal P}_\omega^* > 0 $, with the
threshold value given according to Eq.~(\ref{result}) by
${\cal P}_\omega^*=
\left(4\delta^2\frac{\pi\omega}{\omega_c}\sin
\frac{2\pi\omega}{\omega_c}
-\sin^2\frac{\pi\omega}{\omega_c}\right)^{-1}$.
The evolution of a $B$ dependence of the photoresistivity $\rho_{\rm ph}$
with increasing microwave power ${\cal P}_\omega^{(0)}={\cal
P}_\omega\:(\omega_c=0)$ is illustrated in Fig.~\ref{fig1}.

\begin{figure}
 \narrowtext
 \centerline{ {\epsfxsize=7cm{\epsfbox{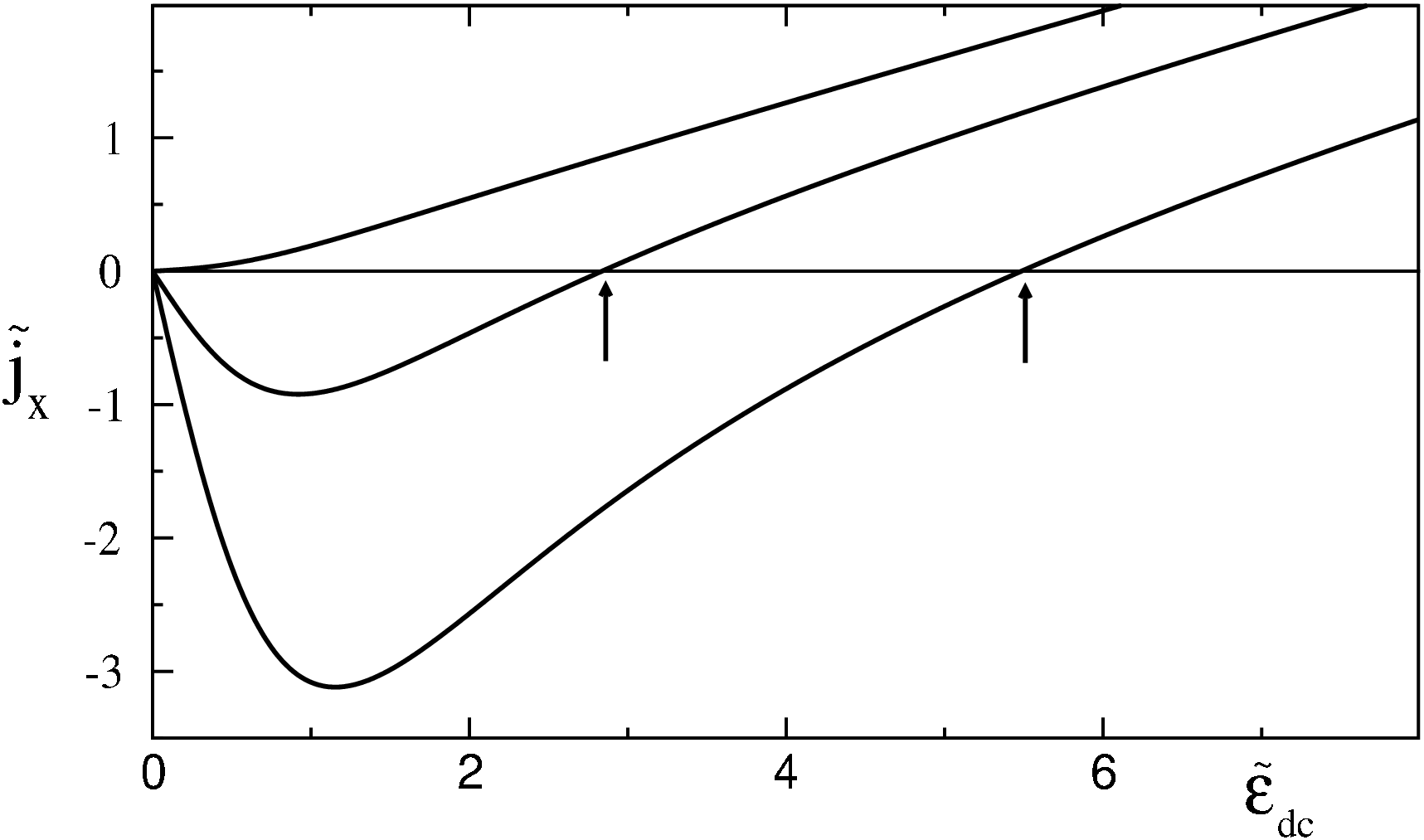}} }}
 \vspace{3mm}
 \caption{Current--voltage characteristics [\,dimensionless current
$\tilde{j}_x=(\sigma_{\rm ph}/\sigma^{\rm D}_{\rm dc}) \tilde{\cal
E}_{\rm dc}$ vs dimensionless field $\tilde{\cal E}_{\rm dc}={\cal
Q}_{\rm dc}^{1/2}$\,] at the points of minima marked by the circles in
Fig.~1. The arrows show the {\it dc} field $\tilde{\cal E}_{\rm dc}^*$
in spontaneously formed domains.}
\label{fig2}
 \end{figure}

Let us now fix $\omega/\omega_c$ such that ${\cal P}_\omega^*>0$,
and consider the dependence
of $\sigma_{\rm ph}$ on the {\it dc} field ${\cal E}_{\rm dc}$ at
${\cal P}_\omega>{\cal P}_\omega^*$. As
follows from Eq.~(\ref{result}), in the limit of large ${\cal E}_{\rm
dc}$ the conductivity is close to the Drude value and thus positive, 
$\sigma_{\rm ph}=(1-6\delta^2)\sigma_{\rm dc}^{\rm D} >
0$. Therefore, $\sigma_{\rm ph}$ changes sign at a certain value
${\cal E}_{\rm dc}^*$ of the {\it dc} field, which is determined by
the condition ${\cal Q}_{\rm dc}=
({\cal P}_\omega-{\cal P}_\omega^*)/{\cal P}_\omega^*$,
see Fig.~\ref{fig2}. The
negative-conductivity state at ${\cal E}_{\rm dc}<{\cal E}_{\rm dc}^*$
is unstable with respect to the formation of domains with a spontaneous
electric field of the magnitude ${\cal E}_{\rm dc}^*$ \cite{andreev03}.

Using Eqs.~(\ref{unitsPQ}), we obtain
\bea
\label{e-domain}
{\cal E}_{\rm dc}^*&=&
\sqrt{{\cal E}_\omega^2-({\cal E}_\omega^*)^2}
\left[\frac{\omega_c^4 (\omega^2+\omega_c^2)}{2
\omega^2(\omega^2-\omega_c^2)^2}\right]^{1/2}
\\
&\times&
{1\over\pi}{\rm Re}\left(4\delta^2\frac{\pi\omega}{\omega_c}\sin
\frac{2\pi\omega}{\omega_c}
-\sin^2\frac{\pi\omega}{\omega_c}\right)^{1/2},
\nonumber
\eea
with ${\cal E}_\omega^*$ being the threshold value of the {\it ac}
field at which the zero-resistance state develops.
Equation (\ref{e-domain}) relates the electric field formed in the
domain (measurable by local probe \cite{willett03})
with the excess power of microwave
radiation. It is worth noticing that this relation does not include
the rate of the inelastic processes.

\begin{figure}
 \narrowtext
 \centerline{ {\epsfxsize=7cm{\epsfbox{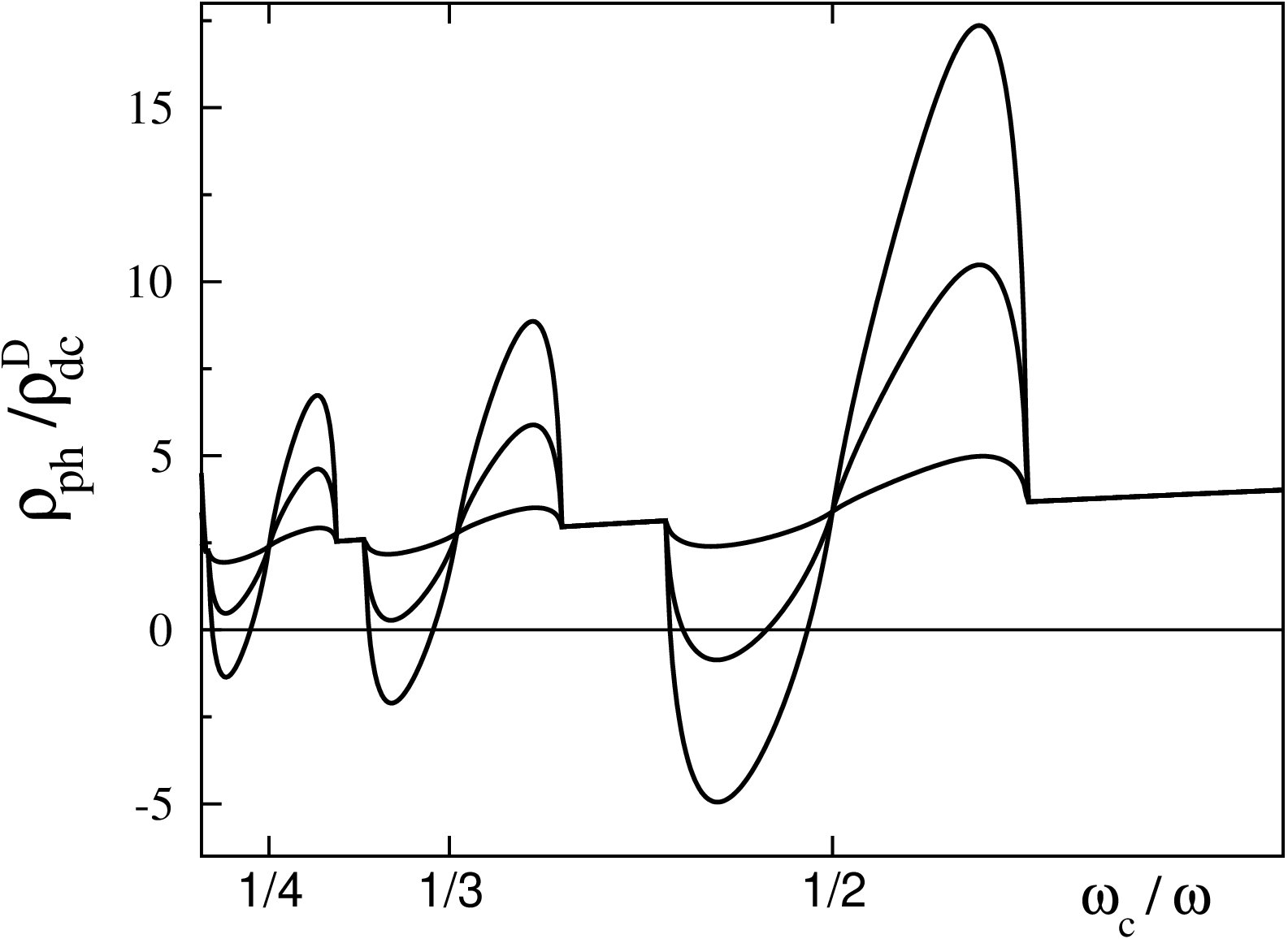}} }}
 \vspace{3mm}
 \caption{Photoresistivity (normalized to the dark Drude value) for
separated Landau levels vs $\omega_c/\omega$ at fixed
$\omega\tau_{\rm q}=16\pi$. The curves correspond to
different levels of microwave power ${\cal P}_\omega^{(0)}
=\{0.004,\,0.02,\,0.04\}$. }
 \label{fig3}
 \end{figure}

We now turn to the regime of strong $B$, $\omega_c\tau_{{\rm q}}/\pi\gg
1$, where the LLs get separated. The DOS is then given (within the
self-consistent Born approximation) by a sequence of semicircles of
width $2\Gamma=2(2\omega_c/\pi\tau_{{\rm q}})^{1/2}$:
\be
\label{SepDOS}
\tilde{\nu}(\ve)=\frac{2\omega_c}{\pi\Gamma^2}
\sum_n {\rm Re}\,
\sqrt{\Gamma^2-\left(\ve-n\omega_c-\omega_c/2\right)^2}.
\ee
We use Eqs.~(\ref{photo}) and (\ref{kineq}) to evaluate the
OPC at ${\cal Q}_{dc}\to 0$ to  first order in
$\cal P_\omega$ and estimate the correction of the second order.
We obtain
\bea 
\label{1order}
&&{\sigma_{\rm ph}\over\sigma^{\rm D}_{\rm dc}}
={16\omega_c\over3\pi^2\Gamma}  
\left\{1-{\cal P}_\omega {\omega\omega_c\over\Gamma^2} \right. \nonumber \\
&& \phantom{aaa} \times \left.\left[
\sum_n \Phi\left({\omega-n\omega_c\over\Gamma}\right)
+ O\left({\omega_c {\cal P}_\omega\over\Gamma}\right)
\right] \right\}, \\
&&\Phi(x)\!=\!\frac{3 x}{4\pi}{\rm Re}\!\left[ {\rm arccos}(|x|-1)-
{1-|x|\over 3}\sqrt{|x|(2-|x|)} \right].
\nonumber
\eea
The photoresitivity for the case of
separated LLs, Eq.~(\ref{1order}), is shown in
Fig.~\ref{fig3} for several values ${\cal P}_\omega$ of the microwave power.
Notice that a correction to Eq.~(\ref{1order}) of second
order in ${\cal P}_\omega$ is still
small even at ${\cal P}_\omega > {\cal
  P}_\omega^*=\Gamma^2/\omega\omega_c$, since
$\omega_c P_\omega^* /\Gamma = \Gamma/\omega \ll 1$.
This means that it suffices to keep the linear-in-${\cal
  P}_\omega$ term
only even for the microwave power at which the linear-response
resistance becomes negative.

As in the case of overlapping LLs, a negative value of the
linear-response conductivity signals an instability leading to the
formation of domains with the field ${\cal E}^*_{\rm dc}$ at which
$\sigma_{\rm ph}({\cal E_{\rm dc}})=0$. It turns out,
however, that for separated LLs the kinetic equation in the form of
Eq.~(\ref{kineq}) yields zero (rather than expected positive)
conductivity in the limit of strong ${\cal E}_{\rm dc}$.
This happens because elastic impurity scattering between 
LLs,  inclined in a strong {\it dc} field, is not included in
Eq.~(\ref{kineq}). The inter-LL transitions become efficient in {\it
dc} fields as strong as ${\cal E}_{\rm dc}^*\simeq (\tau_{\rm
tr}/\tau_{\rm q})^{1/2}\omega_c^2/e v_F$ \cite{vavilov03},
which actually gives the strength of the field in domains.

Finally, we calculate the inelastic relaxation time $\tau_{\rm in}$.
Of particular importance is its $T$ dependence which in turn
determines that of $\sigma_{\rm ph}$. At not too high $T$, the
dominant mechanism of inelastic scattering is due to electron-electron
(e-e) collisions. It is worth emphasizing that the e-e scattering does
not yield relaxation of the total energy of the 2DEG and as such
cannot establish a steady-state {\it dc} photoconductivity. That is to
say the smearing of $f_0(\ve)$ in Eq.~(\ref{fe-expansion}), which is a
measure of the degree of heating, is governed by electron-phonon
scattering. However, the e-e scattering at $T\gg\omega_c$ does lead to
relaxation of the oscillatory term $f_{\rm osc}$ [Eq.~(\ref{distr})]
and thus determines the $T$ behavior of the oscillatory contribution to
$\sigma_{\rm ph}$.

Quantitatively, the effect of electron-electron interaction is taken
into account by replacing the right-hand side of Eq.~(\ref{kineq})
by $-{\rm St}_{ee}\left\{ f \right\}$, where the collision integral
${\rm St}_{ee}\left\{ f \right\}$ is given by
\bea
\label{St}
&&{\rm St}_{ee}\left\{ f \right\}=
\int d\ve^\prime\int dE \: A(E)
 \tilde{\nu}(\ve_+)
 \tilde{\nu}(\ve^\prime)\tilde{\nu}(\ve^\prime_-)\\
&&\ \ \times\left[
-f(\ve)f_h(\ve_+)f(\ve^\prime) f_h(\ve^\prime_-)
+  f_h(\ve)f(\ve_+)f_h(\ve^\prime) f(\ve^\prime_-)
\right],   \nonumber
\eea
and $f_h(\ve) \equiv 1-f(\ve)$, $\ve_{+}= \ve + E$,
$\ve_{-}^\prime= \ve^\prime - E$.
The function $A(E)$ describes the dependence of the matrix
element of the screened Coulomb interaction on the transferred energy $E$,
\[
A(E) = \frac{1}{2\pi\epsilon_F}\,
\ln \frac{\epsilon_F}{{\rm max}\left[E,
    \omega_c (\omega_c\tau_{{\rm tr}})^{1/2},
\Gamma (\omega_c\tau_{{\rm tr}})^{1/2}
\right]},
\]
where $\epsilon_F$ is the Fermi energy.
Thus $A(E)$ differs from the corresponding dependence
for a clean 2DEG at zero $B$ only
by a change in the argument of the logarithm (a more detailed discussion
will be given elsewhere).

We linearize the collision integral and solve Eq.~(\ref{kineq}).
For overlapping LLs, we put $\tilde\nu =1$ in accord
with the accuracy of Eq.~(\ref{result}). Then only out-scattering
processes contribute to the relaxation of the oscillatory part of the
distribution function (\ref{distr}); the result is obtained
by replacing $\tau_{\rm in} \to \tau_{ee}(\ve,T)$ in Eq.~(\ref{kineq})
with \cite{note2}
\be\label{overLL}
\frac{1}{\tau_{ee}}=\frac{\pi^2 T^2+\ve^2}
{4 \pi\epsilon_F}\,\ln\frac{\epsilon_F}
{{\rm max}\left[T, \omega_c(\omega_c\tau_{\rm tr})^{1/2}\right]}.
\ee

We turn now to the case of separated LLs.
In this case, due to oscillation of $\tilde\nu$, even the linearized
collision integral gives rise to a non-trivial integral operator.
Analytical solution of the kinetic equation with this collision operator
does not seem feasible. However, up to a factor of
order unity, we can replace the exact collision
integral with the relaxation-time approximation, thus
returning to Eq.~(\ref{kineq}) with
 \be
 \label{sepLL}
 \frac{1}{\tau_{\rm in}}\sim\frac{\omega_c}{\Gamma}\:\frac{
 T^2}{\epsilon_F}\,\ln
\frac{ \epsilon_F}
 {{\rm max}\left[T,
\Gamma \left(\omega_c\tau_{\rm tr}\right)^{1/2}
\right]}.
 \ee
 One sees that in both cases of overlapping and separated LLs the
 inelastic relaxation rate is proportional to $T^2$, so that the
 OPC $\sigma_{\rm ph}-\sigma_{\rm dc}$
 in the linear-in-${\cal P}_\omega$ regime
 [Eqs.~(\ref{linear}), (\ref{1order})] scales as $T^{-2}$.

Our results are in overall agreement with the experimental findings
\cite{mani02,zudov03}. The observed $T$ dependence of the
photoresistivity at maxima compares well with the predicted $T^{-2}$
behavior. Typical parameters $\omega/2\pi\simeq 50-100$~GHz,
$\tau_{\rm q}\simeq 10$~ps yield $\omega\tau_{\rm
q}/2\pi\simeq 0.5-1$ (overlapping LLs), and the experimental data
indeed closely resemble Fig.~\ref{fig1}. For $T\sim 1\:{\rm K}$ and
$\epsilon_F\sim 100\:{\rm K}$ we find $\tau_{\rm in}^{-1}\sim 10\:{\rm mK}$,
much less than $\tau_{\rm q}^{-1}\sim 1\:{\rm K}$, as assumed in our
theory. Finally, for the microwave power $\sim 1$~mW and the sample
area $\sim 1\,{\rm cm}^2$, we estimate the dimensionless power ${\cal
P}_\omega^{(0)}\sim 0.005-0.1$, which agrees with characteristic
values for separated LLs (Fig.~\ref{fig3}) but is noticeably less
than the prediction for overlapping LLs (Fig.~\ref{fig1}). The reason
for this discrepancy remains to be clarified.

To summarize, we have presented a theory of magnetooscillations in the
photoconductivity of a 2DEG. The parametrically largest contribution
to the effect is governed by the microwave-induced change in the
distribution function. We have analyzed the nonlinearity with respect
to both the microwave and {\it dc} fields. The result takes an
especially simple form in the regime of overlapping LLs,
Eq.~(\ref{result}). We have shown that the magnitude of the effect
governed by the inelastic relaxation time increases as $T^{-2}$ with
lowering temperature.

We thank R.R.~Du, K.~von~Klitzing, R.G.~Mani, J.H.~Smet, and
M.A.~Zudov for information about the experiments, and I.V.~Gornyi
for numerous stimulating discussions. This work was supported by the
SPP ``Quanten-Hall-Systeme'' and the SFB195 of the DFG, 
by NSF grants DMR02-37296, EIA02-10376 and
AFOSR grant F49620-01-1-0457, and by the RFBR.

\end{multicols}

\begin{references}
\bibitem{note2} The energy $\ve$ is counted in Eq.~(\ref{overLL}) from
the Fermi level.
The $\ve$ dependence of $\tau_{ee}^{-1}$ in
Eq.~(\ref{overLL}) does not change the $T^{-2}$ scaling (since the
characteristic energy  $\ve$ in Eq.~(\ref{photo})
 is of the order of $T$), but only yields a
numerical factor. In particular, the inelastic time entering Eq.~
(\ref{linear}) is $\tau_{\rm in} = 0.935\, \tau_{ee}(\ve=0,T)$.
\end{references}
\end{document}